\documentclass[letterpaper, 11pt, oneside, notitlepage, twocolumn, ]{article}

\usepackage[margin=0.5in, bottom=0.75in, top=0.5in]{geometry}
\usepackage{graphicx, lipsum, textcomp, wrapfig, amsmath}
\usepackage[utf8]{inputenc}
\usepackage{chemformula}
\usepackage{comment}
\usepackage{subcaption}
\usepackage{qrcode}
\usepackage{xcolor}
\usepackage{float}

\usepackage{titlesec}
\titleformat{\section}
  {\normalfont\Large\bfseries}{\thesection.}{1em}{}
\usepackage{indentfirst}

\usepackage[switch,columnwise]{lineno}

\usepackage{abstract}
\usepackage{titling}

\usepackage{fancyhdr,graphicx,lastpage}
\usepackage[scaled]{helvet}
 
\usepackage[T1]{fontenc}
\usepackage[scaled]{beramono}
\definecolor{darkgreen}{rgb}{0.05, 0.3, 0.1}
\let\oldtexttt\texttt
\renewcommand{\texttt}[1]{\oldtexttt{\textcolor{darkgreen}{#1}}}

\usepackage{tocbasic}
\DeclareTOCStyleEntry[
  beforeskip=.2em plus 1pt,
  pagenumberformat=\textbf
]{tocline}{section}
\DeclareTOCStyleEntry[
  indent=0.5em,
  numwidth=2em
]{tocline}{subsection}

\usepackage{stfloats}
\usepackage[skip=12pt]{caption}
\def \mauro#1 {\textcolor{mauroblue}{#1}}

\usepackage{fontawesome5}
\newcommand{\orcid}[1]{\href{https://orcid.org/#1}{\textcolor[HTML]{A6CE39}{\small \faOrcid}}}

\newcommand{\github}[1]{\href{https://github.com/#1}{\textcolor[HTML]{70726E}{\small \faGithub}}}

\usepackage{hyperref}
\definecolor{mypurple}{RGB}{140,54,140}
\definecolor{homered}{RGB}{127, 0, 10}
\definecolor{officeorange}{RGB}{204, 75, 0}
\definecolor{mauroblue}{RGB}{53, 48, 217}
\definecolor{citegreen}{RGB}{15, 133, 13}
\definecolor{hyperlinkpurple}{RGB}{42, 0, 163}
\hypersetup{
    colorlinks=true,
    linkcolor=hyperlinkpurple,
    filecolor=mypurple,      
    urlcolor=teal,
    citecolor=citegreen
}

\usepackage[
    style=ieee, 
    isbn=false, 
    url=true, 
    natbib=true, 
    backend=bibtex,
    maxcitenames=2,
    mincitenames=1
    ]{biblatex}

\title{
\vspace{-24pt}
\textbf{Data-driven candidate structure ensemble generation beyond the convex hull with \texttt{crystALL} approach}
}

\author{
\textbf{
Adam M. Krajewski~\textsuperscript{a $\star$ $\dagger$}} \orcid{0000-0002-2266-0099} \github{amkrajewski}, 
\textbf{Ricardo Amaral~\textsuperscript{b $\star$}}
\orcid{0000-0003-1070-9964} \github{ricardonpa}, 
Jonathan W. Siegel~\textsuperscript{c}
\orcid{0000-0002-1493-4889},\\
Luke A. Myers~\textsuperscript{a}
\orcid{0009-0003-0823-0871}
\github{lukeamyers},
Rushi Gong~\textsuperscript{a $\ddagger$}
\orcid{0009-0004-0543-581X}
\github{RushiGong},
Shun-Li Shang~\textsuperscript{a}
\orcid{0000-0002-6524-8897},
Zi-Kui Liu~\textsuperscript{a}
\orcid{0000-0003-3346-3696}\\
\footnotesize{
a. Department of Materials Science and Engineering, The Pennsylvania State University, USA}\\
\footnotesize{b.  Department of Energy and Mineral Engineering, The Pennsylvania State University, USA}\\
\footnotesize{c.  Department of Mathematics, Texas A\&M University, USA}\\
\footnotesize{$\star$ equal contributions | $\dagger$ now at Google | $\ddagger$ now at Microsoft}\\
\footnotesize{Corresponding Author: A. M. Krajewski | PO Box 60022, Sunnyvale, CA 94088 | adam@phaseslab.org}
}

\date{\today}
\renewcommand{\maketitlehookd}{%

\begin{abstract}
    Real materials consist of multiple stable or metastable states under real world conditions, existing as statistical ergodic ensembles of configurations rather than ideal crystals. Yet, data-driven computational materials discovery increasingly biases itself toward finding 'new stable structures' based \emph{only} on DFT ground-state calculations. 
    This can limit our long-term ability to discover materials whose properties emerge from metastable phases, explain experimental observations, or model thermodynamic behavior comprehensively. Predicting ergodic sets of atomic configurations of existing and hypothetical solids is a fundamental challenge in theoretical crystallography, approached through many complementary methods. 
    We discuss this conceptually and take a step forward by introducing \texttt{crystALL}, a pragmatic and powerful data-driven approach to the discovery of atomic configuration ensembles beyond the convex hull, embracing structural diversity over stability alone, thus enabling robust experiment matching in novel cases evading traditional methods. \texttt{crystALL} leverages millions of relaxed structures from Material-Property-Descriptor Database (\texttt{MPDD}) and \texttt{OPTIMADE}, to rapidly generate a comprehensive set of candidates. 
    It achieves this by: (1) species substitutions; (2) machine learning (ML) feature space analysis with \texttt{pySIPFENN}; (3) representative structure selection through clustering and energy-minimization or property matching; and (4) construction of final candidates ensemble for direct validation against experiments or ab initio calculations. 
    Furthermore, we uniquely address the common experimental challenge where experimental composition measurements carry uncertainty spanning many integer stoichiometries, introducing novel and efficient searches across anonymized compositional ranges rather than fixed formulas. Overall, \texttt{crystALL}'s ensemble-focused philosophy provides critical advantages over the prevailing convex-hull-structures-searching paradigm, which often cannot account for or miss important experimental observations and increases AI/ML discovery bias over time - akin to model collapse seen in Large Language Models (LLMs) when retrained on iteratively generated content.
    
    \rule{0.92\textwidth}{0.5pt}
\end{abstract}
}

\begin{document}

\maketitle

\tableofcontents 


\section{New Stable Structure Prediction?} \label{sec:newstable}

The ability to determine the low-energy atomic configurations of existing and hypothetical solids formed in chemical systems is one of the fundamental challenges in theoretical crystallography and in materials science as a whole. For the last several decades, the two remaining challenges are predictions of (a) compositions forming compounds of interest and (b) candidate topologies or approximate atomic arrangements they can form \cite{Hawthorne1994StructuralCrystals}. Given these two, the exact structure can be solved experimentally, by matching X-ray diffraction profile to candidates using the method by \citet{LeBail1988Ab-initioDiffraction} and interpreting the profile, or more recently using deep learning \cite{Yue2024PhaseLearning, Mehdi20252D-diffractogramGeneration, Qin2025GCN-BasedIdentification} including initial crystallographic indexing via machine learning (ML) methods \cite{Shu2025MachineSystems}. Similar analysis can be done with different experiments, such as the Raman spectrum analysis \cite{Ferrari2013}. For hypothetical compounds in exactly defined states, the corresponding atomic configuration can also be solved using ab initio quantum mechanical methods, such as Kohn-Sham density functional theory (DFT) \cite{Kohn1965Self-ConsistentEffects}, which are computationally expensive but can provide accurate results and continue to improve \cite{Kothakonda2023TestingCorrection, Sun2015StronglyFunctional}.

The latter problem of proposing the topologies has been traditionally approached by finding geometries satisfying a set of constraints, like the "principles" systematized by \citet{Pauling1929TheCrystals}, which included then-novel concepts like coördination [\textit{sic}]. Over decades, this approach has evolved into automated computational tools like the \texttt{GRINSP} by \citet{LeBail2005InorganicGRINSP} which can propose such structures. More recently, increasing computational power enabled the shift in focus towards testing many options, with some of the methods starting from random arrangements and collapsing them into local minima of the high-dimensional energy landscape, like \texttt{AIRSS} by \citet{Pickard2011iAbSearching}, or by evolutionary techniques, like \texttt{USPEX} by \citet{Glass2006USPEXPrediction}, \texttt{CALYPSO} by \citet{Wang2012CALYPSO:Prediction}, or \texttt{XtalOpt} by \citet{Hajinazar2026XtalOptOptimization}.

Several efforts have also been made to tackle both topological and compositional challenges simultaneously, such as high-throughput construction of the \texttt{AFLOW}'s \cite{Curtarolo2012AFLOW:Discovery} DFT Database \cite{Toher2018TheDiscovery, Esters2022Aflow.org:Tools} (and several others \cite{Saal2013MaterialsOQMD, Kirklin2015TheEnergies, Shen2022ReflectionsOQMD, Jain2013Commentary:Innovation, Schmidt2022AFunctionals, Choudhary2020TheDesign}) built by systematically populating the earlier, yet continuously growing, libraries of prototype structures \cite{Eckert2024The4, Anderson2026The5} with new chemistries based on the expert knowledge. However, even for a limited set of elements, attempting to completely explore the materials space becomes combinatorially challenging, if not intractable, due to the vast number of possible atomic configurations these elements can form \cite{Krajewski2024Nimplex}, which is severely amplified by the numerous ways complex structures can be occupied \cite{Krajewski2025EfficientStructures}. This issue has prompted recent efforts into (1) systematic similarity-driven prediction of substitutions which would result in finding new low-energy structures \cite{Wang2021PredictingSimilarity}; (2) large-scale brute-force substitutions into prototypes, then filtered by ML models \cite{Schmidt2023Machine-Learning-AssistedMaterials}; and (3) increasingly more advanced approaches that combine these strategies, such as ones by \citet{Ye2022NovelAgents}, \citet{Merchant2023ScalingDiscovery}, or \citet{Zeni2025ADesign} leading to the rapid discovery of new structures \emph{predicted} to be \emph{stable against a known set} of competing atomic arrangements \emph{at absolute zero and in the absence of any external fields or perturbations}.

Despite these advances in finding new \emph{candidates} for ground-state configurations across many chemical systems, which certainly open the world for new discoveries, key questions remain regarding the inherent limitations of such methods that must be considered when interpreting their results. 

Firstly, \emph{"To what degree have we biased the discovery of new structures?"} through many iterations of biasing the search to our expectations, which is philosophically analogous to how retraining Large Language Models (LLMs) and other ML models on generated data makes them underestimate improbable events causing so-called \textit{Model Collapse} by \citet{Shumailov2024AIData}, and drift away from the distribution of the real data \cite{Alemohammad2023Self-ConsumingMAD}. In the context of materials, the latter is depicted well by Fig. 5 in \citet{Ye2022NovelAgents} showing relatively small overlap between experimentally known structures from ICSD and new ones predicted to be stable - something that can be attributed to either (a) novelty or (b) sampling in regions where there is less energetic competition from known low-energy structures so many iterations of "new stable compounds" can be found with increasing focus. Similar concerns were raised by \citet{Cheetham2024ArtificialDiscovery}, who examined a sample of generatively-predicted "new stable" compounds and found many to be known structure types, trivial substitutions to complete the task, or samples in chemically implausible regions, questioning the novelty and practical value of such predictions driven by the stability alone.

Secondly, even in the case of the results domain becoming strongly biased towards the true solutions of stable configurations, one needs to also ask \emph{"Do we actually need the diverse energy landscapes to discover new materials?"} to which the answer is a multifaceted yes. The metastable materials are critical to evaluating thermodynamic model parameters, as so-called protodata, which was highlighted in the study by \citet{Olson2023GenomicDynamics}. That study has also stressed that computational thermodynamics techniques generally model \emph{phases} which, at finite temperatures, represent statistical ensembles of underlying atomic configurations within a material.

Furthermore, one has to consider that going beyond individual highly-stable configurations is needed to create a realistic model matching experiments, as concisely explained by \citet{Zunger2019BewareMaterials}, because of the several complexities involved in the synthesis of new materials, such as defect modeling or disorder contributions. The latter necessitates accounting for entropic contributions, especially in compositionally complex materials, as shown by \citet{Divilov2024DisorderedDiscovery} using ensembles of metastable partially representative configurations to describe high entropy ceramics and match experiments accurately; or by \citet{Grau-Crespo2025CapturingSolutions} showcasing the need to account for complex compositional fluctuations in otherwise relatively simple materials.

Once such realistic models are obtained, the coexistence of many configurations can be used to accurately (a) tune material properties, including to extreme behavior, such as infinitely negative thermal expansion predicted by \citet{Liu2022ZentropyExpansion} that, critically, matched the experiments without any fitting parameters, and (b) model complex phenomena, such as superconductivity \cite{Ferreira2023SearchSystem}. Thus, to achieve, or at least approximate ergodicity, efforts to generate ensembles of candidate structures must sample diverse populations of polymorphs, such as the one shown in Fig. 6 in \citet{Pickard2011iAbSearching} generated through \texttt{AIRSS} for the simple case of pure silicon.

Combined together, the existing materials discovery approaches mentioned throughout this section cover prediction of (1) focused, narrow sets of likely (to be observed) candidates over singular targeted chemistries based on expert intuition and DFT calculations, such as \texttt{USPEX} \cite{Glass2006USPEXPrediction}, \texttt{AIRSS} \cite{Pickard2011iAbSearching}, or \texttt{XtalOpt} \cite{Hajinazar2026XtalOptOptimization}; (2) narrow sets of \emph{individually} unlikely candidates based on DFT for many chemistries in large-scale mostly unbiased enumerations, like \texttt{AFLOW} \cite{Toher2018TheDiscovery}; (3) narrow sets of individually more likely candidates for many chemistries, although often less likely than expected due to pronounced stability bias, based on ML and high-throughput enumerations \cite{Cheetham2024ArtificialDiscovery}, such as \texttt{GNoME} \cite{Merchant2023ScalingDiscovery}; and (4) wide sets of candidates typically for a single targeted material, considered together and approximately ergodic, created based on the expert intuition and systematic enumeration of configurations, like Zentropy approach by \citet{Liu2022ZentropyExpansion}. These leave out the essential niche of wide sets of candidates for many chemistries, which remains, as mentioned earlier, infeasible to be solved by DFT or experiments alone, yet it can be approached through fully-data-driven DFT/experiment-validated ML.

\section{\texttt{crystALL} Approach} \label{sec:crystall}

Generating a wide set of diverse candidates requires the opposite philosophy  from one that biases the candidate population toward stability based on expert intuition - human or machine alike.
Rather, all possibilities coming from historically realized or relaxed structures should be considered during the initial steps.
In this way, the search is not constrained by prior assumptions about stability, but is instead allowed to reveal unexpected regions of configurational space.

This work presents a concrete approach and associated open-source software tool based in this philosophy, called \texttt{crystALL}, which does so by unbiased consideration of \underline{all} past 3D periodic atomic structures (i.e., \underline{cryst}als) data, including different deviations from ideal positions from DFT-based relaxations in different chemical systems, in an attempt to accommodate targeted composition or its range into it. This is akin to populating prototypes from a library like the \texttt{AFLOW} \textit{Library of Crystallographic Prototypes} \cite{Eckert2024The4, Anderson2026The5}; however, instead of a highly curated set of around 2,000 (as of March 2026) prototypes with ideal sites and commonly occurring crystallographic parameters, three orders of magnitude more prototype atomic configurations are used, including denser sampling of crystallographic free parameters determining ideal site positions, as well as symmetry-broken configurations, which have been metastable for other chemistries.

At its core, \texttt{crystALL} first connects to an expansive collection of configurations in the \texttt{MPDD} \cite{Krajewski2021MPDD:Database, Evans2024DevelopmentsExchange}, developed by the authors with this application in mind, which spans over 4.9 million \emph{relaxed} structures (as of March 2026) coming from (1) experimental observations processed through Crystallography Open Database (COD) developed by \citet{Grazulis2009CrystallographyStructures, Grazulis2012CrystallographyCollaboration, Grazulis2019CrystallographyPerspectives}, and (2) various DFT calculations through OQMD \cite{Saal2013MaterialsOQMD, Kirklin2015TheEnergies, Shen2022ReflectionsOQMD}, AFLOW \cite{Curtarolo2012AFLOW:Discovery, Toher2018TheDiscovery, Esters2022Aflow.org:Tools}, Materials Project \cite{Jain2013Commentary:Innovation, Horton2025AcceleratedProject}, NIST-JARVIS \cite{Choudhary2020TheDesign}, Alexandria \cite{Schmidt2022AFunctionals}, CAMD \cite{Ye2022NovelAgents}, GNoME \cite{Merchant2023ScalingDiscovery}, and in-house DFT calculations done by the authors. All data in MPDD can be accessed both through a limited-access high-performance API and through an open-access OPTIMADE API \cite{Evans2024DevelopmentsExchange} endpoint at \href{http://optimade.mpdd.org/}{optimade.mpdd.org}. Furthermore, \texttt{MPDD} also includes an additional "community" dataset, accessible by \texttt{crystALL}, which also accepts purely hypothetical atomic configurations, providing a pipeline for systematically proposing and sharing unconventional materials, such as superionic phases of different dimensionalities, as in \citet{Liu2026PredictionConditions}, usually not covered by other databases. These configurations were of interest to the authors and different external contributors, but were not calculated through ab initio methods. These include different disordered occupancies (e.g. SQS by \citet{Zunger1990SpecialStructures, vandeWalle2013EfficientStructures}) or hypothetical structurally complex polymorphs. Finally, one can also add prototypes to \texttt{crystALL} locally, after constructing them in any fashion of choice, such as systematic generation of atomic configurations with \texttt{aflow++} based on an AFLOW prototype selected using \texttt{---proto} option adjusted by scanning through ranges of prototype-specific free parameters set by \texttt{---params} option \cite{Oses2023Aflow++:Design}.

\subsection{Searching for Experimentally Observed Target Stoichiometries} \label{ssec:compounds}

\begin{figure*}[b]
    \centering
    \includegraphics[width=0.875\textwidth]{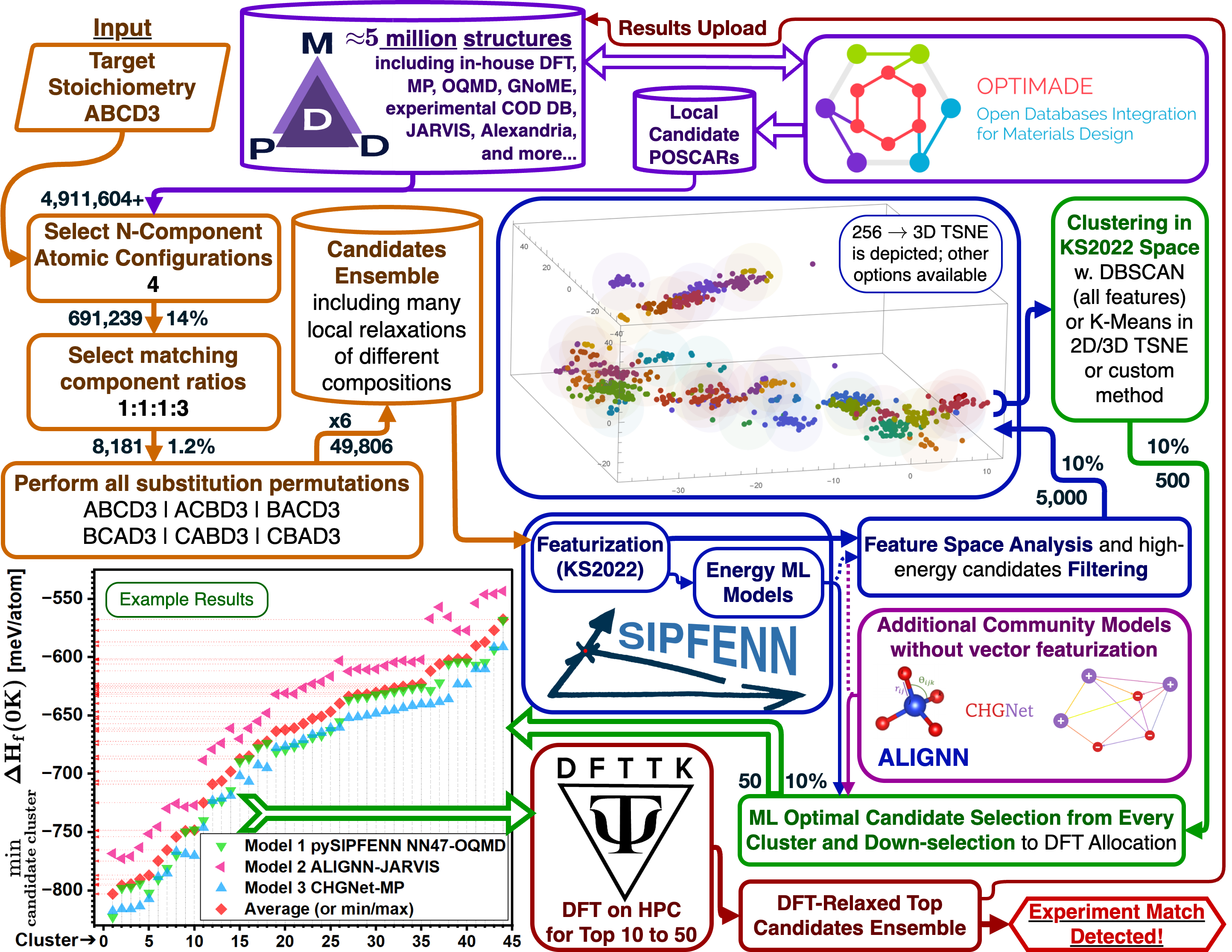}
    \caption{Core schematic of the \texttt{crystALL} approach over a specific stoichiometry, illustrating the complete workflow from candidate generation (orange), featurization and ML with \texttt{pySIPFENN} \cite{Krajewski2025EfficientStructures} (blue),  addition of community ML models (purple), clustering to select top candidates from unique structures/energy basins (green), and final DFT relaxations leading to match with experimental observations (red). The example depicts a dramatic filtering cascade - 5M structures from \texttt{MPDD} are harnessed to generate around 50k \ch{ABCD3} candidates, then ML-filtered to 5k in 500 clusters, culminating in DFT calculations for top 50 cluster representatives to construct an approximately ergodic ensemble and match experimental observations; while feedback loop to \texttt{MPDD} creates a self-improving ecosystem for materials discovery that grows with \texttt{crystALL} use.}
    \label{fig:crystallcompound}
\end{figure*}

The combined pool of diverse atomic configurations provides the starting-point \emph{prototypes} for the rest of the \texttt{crystALL} implementation, depicted schematically in \autoref{fig:crystallcompound} for compounds with a defined stoichiometry and expanded upon later in \autoref{ssec:compositionranges} to cover compositional ranges. 

Given a target chemical composition/stoichiometry, which can be anonymized to, for instance, \ch{ABCD3}, \texttt{crystALL} first filters incoming \texttt{MPDD} data to match it by selecting the appropriate number of components (i.e., $4$) and component ratios (i.e., 1:1:1:3). Then, using atomic structure representations and manipulation routines from \texttt{pymatgen} \cite{George2026Pymatgen:Prospects}, it generates \emph{all} possible permutations of how atoms in the prototype can be substituted with target chemical elements. Depending on the ratios, this can be just 1 for compounds with non-repeating ratios (e.g., in Sec. \ref{ssec:ndbi2}), 2 for a single matching ratio (e.g., in Sec. \ref{sec:novelhydrides}), 6 for the \ch{ABCD3} example, highlighted in the orange-colored part of \autoref{fig:crystallcompound}, or many more, such as 48 permutations possible in complex compounds like \ch{Ba2CaCeGd2Ti2(CuO7)2} (mp-1229095). At this stage, prototypes become an ensemble of \emph{candidates} that can be passed into AI/ML and other materials informatics frameworks.

In addition to restricting searches by target anonymized composition,
users may refine the prototype pool
using any queryable field within \texttt{MPDD},
including dimensionality, crystal system, or any combination of MPDD fields.
For example,
\texttt{crystALL} may be used with \texttt{MPDD} to identify two-dimensional prototypes
with a given composition,
providing a route to construct wide and diverse datasets
of two-dimensional materials.
\texttt{MPDD} supports this functionality using two popular algorithms by \citet{Cheon2017DataHeterostructures} and by \citet{Larsen2019DefinitionComponents}, which assign a label (\texttt{0D} / \texttt{1D} / \texttt{2D} / \texttt{3D}) corresponding to the identified  dimensionality of material (Cheon) or an array of labels identifying dimensionality of all features (Larsen). \texttt{MPDD} provides a field combining these properties and an agreement score that quantifies confidence across the methods at various settings, but users interested in mixed-dimensional structures
(e.g., layered materials with intercalated ions)
can query the raw per-algorithm dimensionality fields directly.
As of writing of this work,
\texttt{MPDD} contains approximately
4.2 million 3D, 430{,}000 2D, 50{,}000 1D, and 20{,}000 0D entries.
Although dimensionality is emphasized here,
the same filtering functionality applies to any combination of queryable fields in \texttt{MPDD}.

These prototypical configurations, handled as \texttt{pymatgen} \cite{George2026Pymatgen:Prospects} \texttt{Structure} objects, are then passed to the open-source \texttt{pySIPFENN} (python toolset for Structure-Informed Property and Feature Engineering with Neural Networks) \cite{Krajewski2022ExtensibleNetworks} where they are featurized into high-dimensional vectors (256D) in \texttt{pySIPFENN}'s \texttt{KS2022} space \cite{Krajewski2025EfficientStructures}, as of the current \texttt{crystALL} version, which is a computationally and scientifically improved version of \texttt{Ward2017} pure-Python space featurizer from earlier \texttt{pySIPFENN} versions. Both were implemented based on statistics over global structure features and local chemical environments building upon the approach first introduced in \texttt{Magpie} by \citet{Ward2017IncludingTessellations}. Such a description of each candidate configuration captures both structural and compositional information, which can be used to analyze the large pool of candidates in terms of uniqueness and to represent different basins in the energy landscape.

The recognition of similar configurations of the same structure energy basins is accomplished through different strategies of clustering candidates in the \texttt{KS2022} feature space, depicted in blue-colored part of \autoref{fig:crystallcompound}. In the current \texttt{crystALL} version, this can be done either (a) across all dimensions encoding structural information, scaled based on min/max values, using DBSCAN (Density-Based Spatial Clustering of Applications with Noise) \cite{Ester1996ANoise}, or (b) via a two-step process of non-linear dimensionality reduction into 2D or 3D (e.g. Fig. \ref{fig:ndbi2clusters}) through t-distributed Stochastic Neighbor Embedding (tSNE) \cite{Hinton2002StochasticEmbedding, VanDerMaaten2008VisualizingT-SNE} followed by $k$-means partitioning the data into $k$ clusters, with the number selected by the user based on expert intuition over the candidate pool or through optimization of silhouette scores, allowing the clustering to reflect the natural groupings in the data. The former approach generally allows capturing more complex relationships and pathway information, while the latter approach provides more intuitive visualization and interpretation in the context of where prototypes came from, allowing researchers to adapt the clustering approach to the specific characteristics of their problem. The accompanying software documentation details the tunable parameters for both clustering methods, the rationale for default selections, and practical guidance for users.

Parallel to clustering, each candidate is passed through ML models to predict its energy and other properties of interest. For rapid static property evaluations, \texttt{crystALL} provides built-in access to multi-property models of  \texttt{pySIPFENN} alongside \texttt{MPDD-ALIGNN}, a lightweight, author-maintained fork of \texttt{ALIGNN} \cite{Choudhary2021AtomisticPredictions} optimized for deployment over \texttt{MPDD}, which includes a default set of individual property models based on the Materials Project \cite{Jain2013Commentary:Innovation, Horton2025AcceleratedProject}, NIST-JARVIS \cite{Choudhary2020TheDesign}, and Alexandria \cite{Schmidt2022AFunctionals}. 

Furthermore, in the upcoming \texttt{v0.6.0} release (currently available as an active development branch on GitHub; see \autoref{sec:softwareavailability}), users can utilize the \texttt{ASE} (Atomic Simulation Environment) \cite{HjorthLarsen2017TheAtoms} interface, allowing users to perform ML-based relaxation of candidate structures utilizing machine-learned interatomic potentials (MLIP), such as \texttt{CHGNet} \cite{Deng2023CHGNetModelling}, \texttt{MACE} \cite{Batatia2025ThePotentials, batatia2022mace}, \texttt{Orb-v3} \cite{Rhodes2025Orb-v3:Scale}, \texttt{MatterSim} \cite{Yang2024MatterSim:Pressures}, or \texttt{UPET} \cite{Malosso2026High-qualityLearning, bigi2026upet}. \autoref{app:mlips} provides further discussion on the 24 available MLIPs. \texttt{crystALL}'s extensibility stems from not only the seamless incorporation of any state-of-the-art model compatible with \texttt{ASE}, \texttt{ALIGNN}, or \texttt{pySIPFENN} via a simple change of config files, but also architectural choices allowing for easy code changes to incorporate any custom model the user may want to include. At the end of the ML pipeline of \texttt{crystALL}, all predictions are combined and can be used individually or based on min/max/mean statistics. This multi-objective capability provides flexibility and convenience to test different model implementations, while allowing for a more nuanced analysis considering other specified physical properties beyond just stability.

Then, using the ML modeling and clustering results, \texttt{crystALL} selects the \textit{top candidate} configuration from each cluster based on user-selected optimality metric, as depicted in the green-colored part of Figure~\ref{fig:crystallcompound}. Optimality may correspond to lowest predicted energy, closest match to an experimentally observed property (for example band gap), or a composite objective defined symbolically based on multiple ML models. For example, one can effortlessly pass an equation like $\sqrt[3]{3\big/\big((M_A/V_0)^{3/2}\big[(B+\tfrac{4}{3}G)^{-3/2}+2G^{-3/2}\big]\big)}$ to compute polycrystalline speed of sound $v_m$ from known mass per atom $M_A$ and ML-predicted shear modulus $G$, bulk modulus $B$, and volume per atom $V_0$. In a typical use case, such as those discussed in \autoref{sec:casestudies}, optimality corresponds to the lowest energy in each cluster. Thus, top candidates can be interpreted as structurally diverse atomic configurations at the local minima of respective ML energy basins within a high-dimensional energy landscape spanning beyond the initial convex hull, often both above and below.

Finally, for materials in well-explored regions of the chemical space, where the accuracy of ML models is reliably high, the \textit{ensemble of top candidates} can be used directly. For novel or complex regions, the candidates are passed to a final validation step using DFT \cite{Kohn1965Self-ConsistentEffects, Kothakonda2023TestingCorrection, Sun2015StronglyFunctional}, which can be executed outside of \texttt{crystALL} via integrated tools like \texttt{DFTTK} \cite{Hew2025DensityApproximation} on high-performance computing clusters. The latter is typically performed in batches on top 10 to 200 configurations until either (1) an approximately ergodic ensemble is constructed, in the case of materials discovery use, or (2) a match or matches (see Sec. \ref{ssec:ndbi2}) to the experimental observations are found, as depicted in red-colored part of \autoref{fig:crystallcompound}. Importantly, the new DFT-calculated structures are contributed back into \texttt{MPDD} allowing their re-use in next iterations significantly improving future results over similar compounds.

\subsection{Searching for Composition Ranges} \label{ssec:compositionranges}

To recognize limitations of single-composition-targeting \texttt{crystALL} searches, discussed in terms of implementation in \autoref{ssec:compounds} and examples later in \autoref{sec:casestudies}, one has to consider consequences of their operation on exact chemical formulas, also called stoichiometries. These can be set precisely, based on (a) highly accurate and precise experimental composition measurements; (b) sufficient composition measurements, combined with expert intuition; or (c) simple assumption that something may be of interest, e.g., enough to be screened \cite{Curtarolo2012AFLOW:Discovery, Schmidt2023Machine-Learning-AssistedMaterials, Ye2022NovelAgents}. Additionally, the exact/nominal nature of compounds, allows researchers to access atomistic data on them readily from nearly any atomistic database thanks to the \texttt{OPTIMADE} API \cite{Evans2024DevelopmentsExchange}. 

\begin{figure*}[b]
    \centering
    \includegraphics[width=0.95\textwidth]{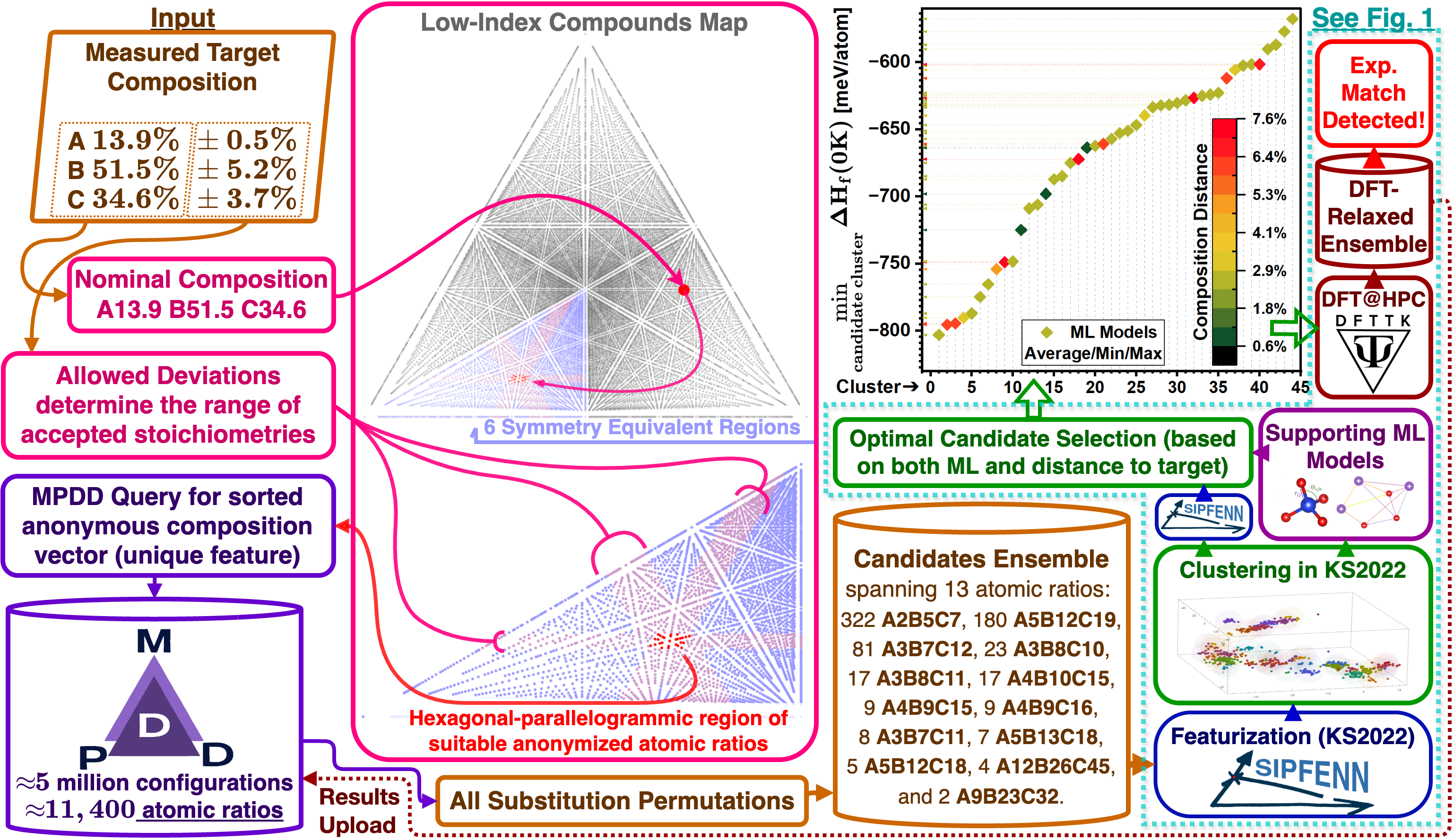}
    \caption{Core schematic of the \texttt{crystALL} approach over compositions with defined ranges, illustrating the complete workflow from definition of the search query (pink) and multi-composition candidate ensemble generation (orange), and further steps described in \autoref{fig:crystallcompound}. The example depicts a composition is matched to 13 compounds within the defined bounds and then analyzed both in terms of stability and distance.}
    \label{fig:crystallcomposition}
\end{figure*}

However, this apparent precision masks a fundamental challenge to purely data-driven structure identification and discovery under point \emph{b}. Namely, the vast majority of the experimental observations do not provide unambiguous stoichiometry due to both: (1) mean/nominal compositions never exactly matching any "low-index" (small integer ratios) compound atomic ratios like \ch{AB3}, \ch{A2B7}, or \ch{A2BC3D3}; and (2) the measurement uncertainty bounds spanning many of such ratios.

This compositional ambiguity can often be overcome with educated guesses based on similar elements and sets of rules, or extensive ab initio efforts until a match is found; however, there are at least two cases where these often fail. First case covers compounds with uncommon or complex structures, such as one called \ch{Mg32Al47Cu7} based on its approximate composition which has been reported for years as the minority phase in several aerospace Al alloys \cite{Li2005Localized6.1} and more recently in lightweight high entropy alloys (L-HEAs) \cite{Sadeghi2022DesignLMLW-HEA}. These typically have a known XRD pattern but no corresponding crystal structure with a well-defined stoichiometry. The second case represents an epistemic challenge underexamined by the community - when exploratory studies investigate uncommon chemical systems without crystallographic measurements and report observing compounds of nominal compositions which are later picked by computational researchers who propose some candidates, identify lowest-energy ones, but conclude that the compound is metastable. Such assessment may be correct, but cannot be proven. To illustrate this epistemic trap, consider a hypothetical first-time study of the \ch{Al-Cu} system where Energy Dispersive Spectroscopy (EDS) yields an \ch{Al_{32.1}Cu_{67.9}} composition. Based on this, one could calculate many candidates for \ch{AlCu2}, which are all metastable, without ever considering \ch{Al4Cu9} even though it is equally distant in the compositional space from the EDS result and actually stable in a $\gamma$-brass-like structure.

The preceding discussion clarifies two critical needs: (a) robust pools of prototypes that span a wide variety of compositions, extending the concept in \autoref{fig:crystallcompound}; and (b) the removal of bias toward only "low-index" compositions, which allows harvesting of all available community data (i.e., approximately 11,393 distinct stoichiometric ratios in \texttt{MPDD} as of March 2026). To address this need, \texttt{crystALL} enables searches for compositional ranges, where the user specifies both the nominal/mean composition and uncertainty in all of the components, individually, as shown in the top left of  \autoref{fig:crystallcomposition}.

To enable this kind of approach, one small yet elegant innovation had to be implemented within \texttt{crystALL} and \texttt{MPDD}. That's because, while most atomistic datasets and \texttt{OPTIMADE} API \cite{Evans2024DevelopmentsExchange} include both anonymized chemical formulas and dictionaries of element fractions, these only allow queries for limited predefined sets or for fraction bounds over specific elements. To assure completeness, the former would require the user to fetch all anonymous formulas every time and filter them for a given bound. The latter can, in principle, work when searching for 2/3-component prototypes, but fails for higher order systems, like 5-component ones, where one has to set $N \cdot N! \cdot 2$, i.e. $5 \cdot 5! \cdot 2 = 1,200$ bounds for each filtered chemical system to look for range of anonymized compositions. In \texttt{crystALL}, this issue has been addressed by including a sorted list of atomic fractions in \texttt{MPDD}, under the \texttt{material.sortedFractionsDict} field. This construction enables both (a) single queries to cover all N-component systems and (b) the collapse of symmetrically-equivalent regions of the composition space $N!$-fold, as depicted in the pink-colored part of \autoref{fig:crystallcomposition}; thus, making queries feasible for any system complexity. To put it in perspective, querying for a modest 5-component composition range over all 84,000+ quinary systems in \texttt{MPDD}, requires over 10,000,000 times fewer bounds set compared to using the typical dictionary of defined elements and their fractions.

Once all atomic ratios within set bounds and corresponding prototype configurations are obtained, they are processed through \texttt{crystALL}'s atomic substitution steps of permutations, featurization, clustering, ML modeling, candidate selection, and, optionally, ab initio steps (e.g., DFT/DFTTK); in the same fashion as described in Section~\ref{ssec:compounds}, with two additional considerations. Firstly, since there are variations in candidate compositions, formation energies alone can no longer be used for direct stability comparisons between them. Thus, all candidates are referenced to the self-consistent ML-computed energy convex hull from \texttt{MPDD} either yielding stability values that can be analyzed directly or allowing for the introduction of a composition-dependent correction factor based on the hull depth at each stoichiometry. Secondly, such an approach generates output with additional dimension of the compositional distance to the nominal composition value, as shown in the example chart in \autoref{fig:crystallcomposition}, enabling researchers or AI agents to make a better-informed decision on what to pass into the validation steps, balancing the predicted stability with closeness to the experimental match.

\section{Case Studies} \label{sec:casestudies}

\subsection{Identifying \ch{NdBi2} and Other Structures} \label{ssec:ndbi2}

The \texttt{crystALL} approach was first deployed by the authors to address an unsolved materials challenge within the study by \citet{Im2022ThermodynamicModeling}. That study reassessed thermodynamic models for the \ch{Nd-Bi} chemical system, representing an essential step towards the rational design of rare-earth alloys for clean energy applications, such as the electrochemical recovery of rare-earth elements and liquid metal batteries. As highlighted by \citet{Im2022ThermodynamicModeling}, while the \ch{NdBi2} compound has long been known to be thermodynamically stable \cite{Yoshihara1975Rare-earthBismuthides}, its crystal structure has remained unknown for five decades, posing a significant obstacle to ab initio investigations. By deploying \texttt{crystALL}'s data-driven approach on \texttt{MPDD}, an ensemble of candidate configurations for \ch{NdBi2} was generated and later validated by DFT-based methods, following an earlier development version of the workflow depicted in \autoref{fig:crystallcompound}.

\begin{figure}[h]
    \centering
    \includegraphics[width=0.425\textwidth]{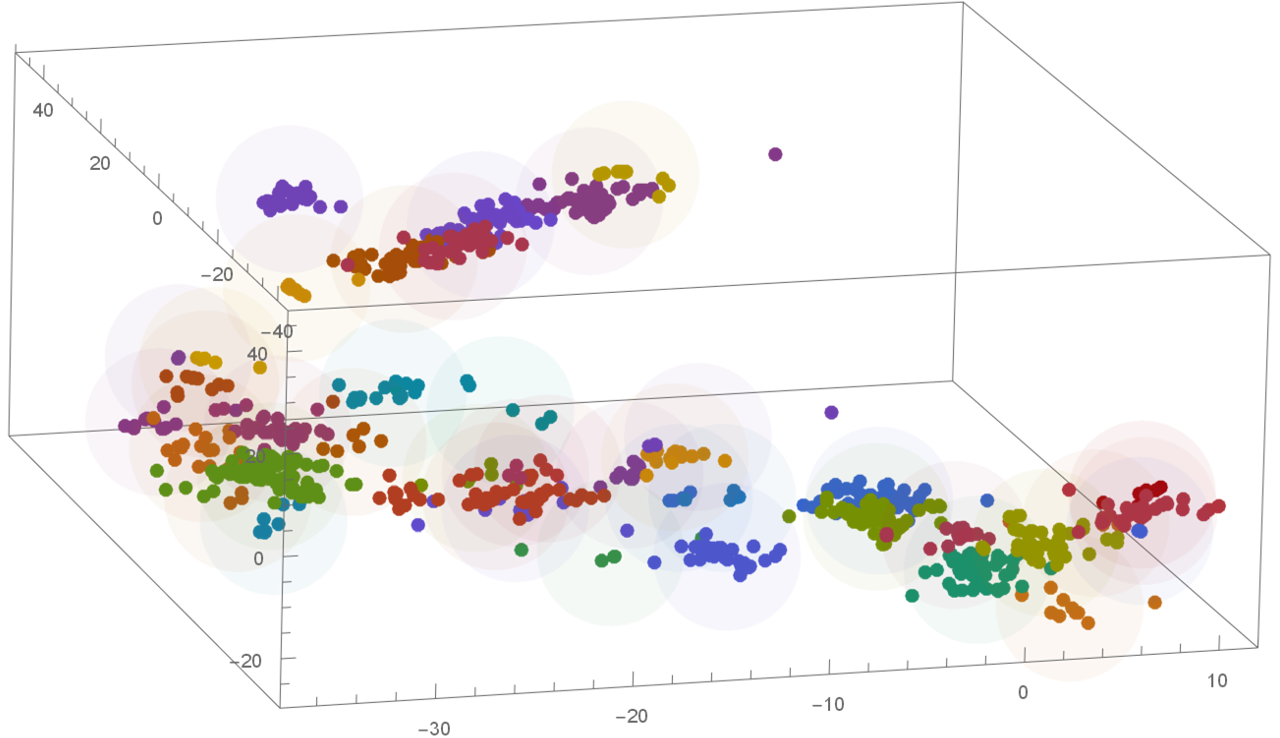}
    \vspace{-6pt}
    \caption{Visualization of \texttt{crystALL}'s tSNE low-dimensional (3D) embedding of \texttt{Ward2017} feature vectors \cite{Krajewski2022ExtensibleNetworks, Ward2017IncludingTessellations} of 1,000 lowest energy \ch{NdBi2} candidates, as predicted by \texttt{SIPFENN}, clustered with k-means approach to determine the ensemble of unique basins.}
    \label{fig:ndbi2clusters}
\end{figure}

First, 26,055 \ch{AB2} prototype configurations were extracted from \texttt{MPDD}, which at the time had approximately 1.3 million DFT-relaxed or experimentally observed configurations covering all structural records contained in OQMD \cite{Kirklin2015TheEnergies}, Materials Project \cite{Jain2013Commentary:Innovation, Horton2025AcceleratedProject}, NIST's JARVIS \cite{Choudhary2020TheDesign}, and the Crystallography Open Database (COD) \cite{Grazulis2019CrystallographyPerspectives}. Following the extraction and substitution, all generated candidates were featurized using the \texttt{Ward2017} featurizer in \texttt{pySIPFENN}, and their energies were predicted via the \texttt{SIPFENN-NN20} "Novel Materials Model" optimized for extrapolations \cite{Krajewski2022ExtensibleNetworks}. The top 1,000 lowest-energy candidates ($\simeq4\%$) were then selected and clustered using k-means (following the 3D tSNE approach described in \autoref{ssec:compounds}) into an intentionally high number of clusters to capture smaller variations, as depicted in \autoref{fig:ndbi2clusters}.

Next, the \texttt{SIPFENN-NN20}-predicted formation energies were used to select the most stable candidate from each cluster, which were each relaxed using three different DFT exchange-correlation functionals (PBE \cite{Perdew1996GeneralizedSimple}, PS/PBESol \cite{Perdew2008RestoringSurfaces}, and SCAN \cite{Sun2015StronglyFunctional}) to assess their order of predicted stability. For comparison, an alternative evolutionary structure search was performed using \texttt{USPEX} by \citet{Lyakhov2013NewUSPEX}, discussed earlier in \autoref{sec:newstable}.

\begin{figure}[h]
    \centering
    \includegraphics[width=0.35\textwidth]{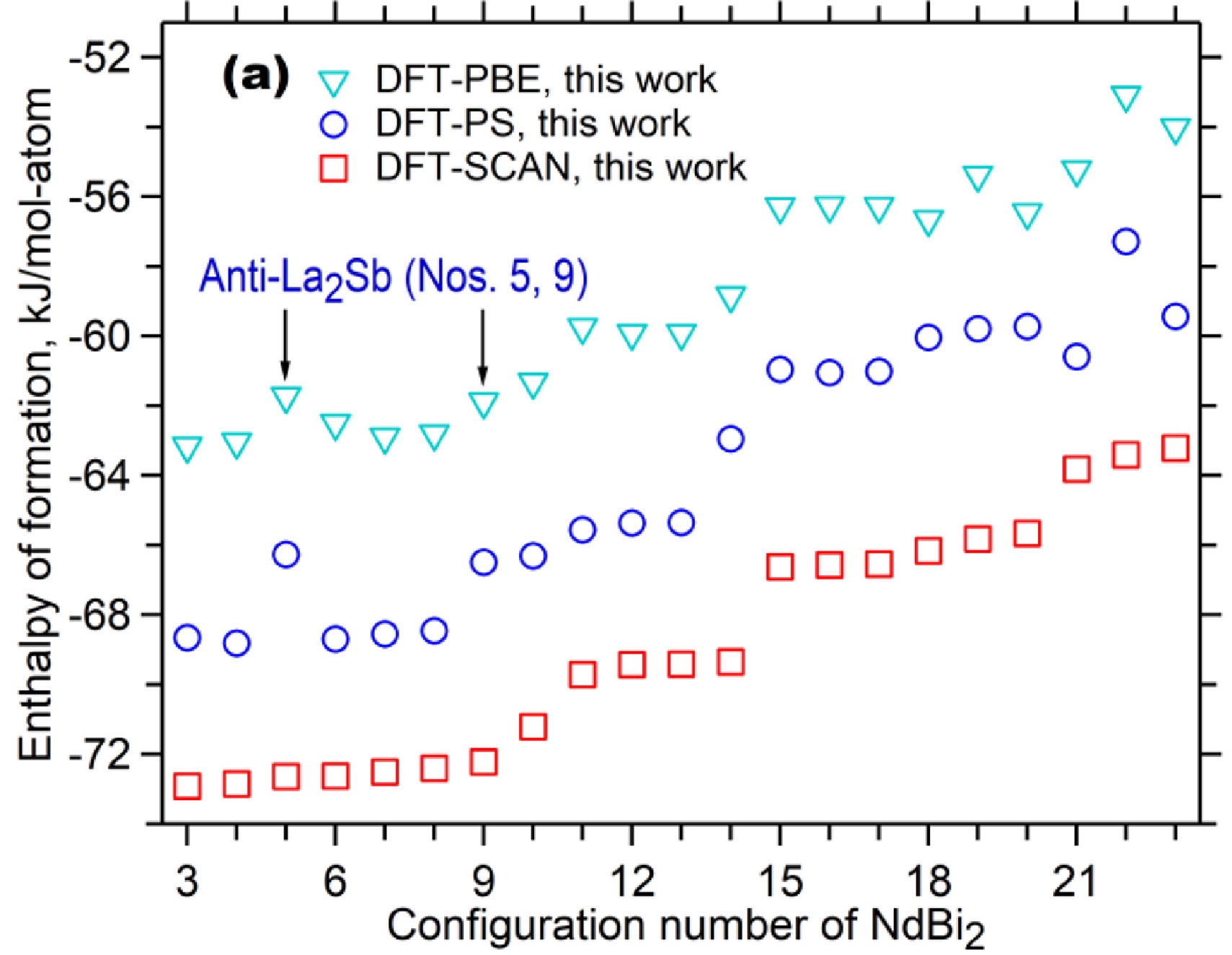}
    \caption{0K formation energies of structure candidates predicted by \texttt{USPEX} (\#3) and \texttt{crystALL} (\#4-23). Pure element reference configurations \#1 and \#2 are omitted. The mixture of 2nd and 6th lowest-energy candidates has been matched to the experimental observations of \ch{NdBi2} demonstrating both \texttt{crystALL}'s predictive ability and importance of studying of ensembles of metastable structures. Reproduced under license from \citet{Im2022ThermodynamicModeling}.}
    \label{fig:ndbi2dft}
\end{figure}

The DFT results, shown in Figure \ref{fig:ndbi2dft} (reproduced from \citet{Im2022ThermodynamicModeling}), indicated that \texttt{USPEX} indeed found the stable structure under SCAN and PBE functionals, relative to all of the 20 structures generated by \texttt{crystALL}. However, as explained in \autoref{sec:newstable}, the zero-temperature lowest energy structure is not necessarily the one experimentally observed. This has been precisely the case in \citet{Im2022ThermodynamicModeling}, as the XRD pattern did not match the \texttt{USPEX}-generated structure. Instead, it was successfully matched to a mixture of two related configurations (anti-\ch{La_2Sb} and \ch{La_2Te}) generated by \texttt{crystALL}, each predicted to be metastable by all three DFT methods, which are marked by arrows in \autoref{fig:ndbi2dft}.

A similar approach and results were also obtained in a computational study of high-pressure intermetallics formed during welding process in the \ch{Al-Fe} system by \citet{Shang2021FormingJoints}, wherein \texttt{crystALL} has been used to propose a set of candidates for \ch{Al2Fe}, including its ground state configuration, which was further validated via DFT and phonon calculations.

Furthermore, following the successful deployment of \texttt{crystALL} over \ch{Nd-Bi} in \citet{Im2022ThermodynamicModeling}, the compositional range search described in \autoref{ssec:compositionranges}, has been deployed over \ch{Nd-Bi-Sn} chemical system, in a follow-up study that remains unpublished. There, a stable ternary compound has been observed, contrary to the current literature on that system \cite{Xu2017ThermodynamicSystems}, and its structure was rapidly proposed through \texttt{crystALL}.

\subsection{Guided Search of Novel Hydrides} \label{sec:novelhydrides}

Another successful case study of \texttt{crystALL} is demonstrated in the search for novel complex hydrides for \ch{H2} storage, where compositions with higher hydrogen content are extensively researched to enhance volumetric and gravimetric capacities and to match desired structural properties. Within this rationale, specific structures such as perovskites have traditionally been targeted for showcasing improved \ch{H2} adsorption and desorption kinetics associated with their high tolerance to defects, large interstices, and high diffusivity. 

Several perovskite hydrides — such as \ch{CsCaH3} and \ch{NaMgH3} \cite{Park1986, Bouamrane2000, Ikeda2005, Pottmaier2011} — have already been synthesized, and their stability and structures are predicted using the empirical tolerance factor between A and B cations, first introduced by Goldschmidt in 1926 \cite{Goldschmidt2016}. In these examples, structural predictions are based on the idealized spatial arrangement of atoms in the \ch{ABH3} structure, which is often oversimplified for more complex perovskites, such as \ch{KMn(BH4)3} \cite{Schouwink2012, Schouwink2014}, with large ionic groups.

Consequently, crystallographic details of such complex perovskites are still lacking, with only a few experimental cases reported \cite{Ravnsbaek2012, Schouwink2015, Moller2016}, given the challenges associated with their synthesis and the refinement of their structural characterization. In addition, perovskite structures can undergo a series of temperature-induced polymorphic transformations that are of great importance for their emerging physical properties. This sequence of structural transitions is dominated by lattice instabilities, corresponding to ensembles of both octahedral rotations and polar displacements, as well as some degree of orbital-induced effects in most compounds. 

Following the target stoichiometry search methodology in \autoref{ssec:compounds}, the authors investigated the \ch{NaMg(BH4)3} composition system, considering all its stable and metastable candidate structures using \texttt{crystALL}. Appendix \autoref{fig:hydrides_chart} summarizes the 100 most stable candidates, generated based on prototypes from Materials Project \cite{Jain2013Commentary:Innovation, Horton2025AcceleratedProject}, GNoME \cite{Merchant2023ScalingDiscovery}, AFLOW \cite{Esters2022Aflow.org:Tools}, and the Crystallography Open Database (COD) \cite{Grazulis2019CrystallographyPerspectives}, while \autoref{fig:hydrides_cluster} displays their 2D tSNE clustering. The seven clusters identified group together candidates with similar atomic environments, representing primarily hexagonal structures (clusters 1 and 2), triclinic structures (cluster 3), trigonal structures (clusters 4, 5, and 7), and perovskite-type structures (cluster 6). The most stable structure predicted in cluster 6 is based on the \ch{NbV(PO4)3} entry from the GNoME dataset (a9b189292f), and comprises a highly distorted perovskite structure, wherein the A sites are coordinated by 12 \ch{BH4-} anions, with half pointing inwards and half pointing outwards. The corner-sharing coordination exhibited by this prototype represents a previously unobserved borohydride structure, distinct from any previously reported ones, which are characterized by edge-sharing \cite{Schouwink2015} and face-sharing \cite{Moller2016} configurations (see \autoref{fig:hydrides_perovskites}).

The \ch{NaMg(BH4)3} case highlights the significant potential of \texttt{crystALL} not only for the rapid and robust prediction of candidate structures to aid experimental validation, but also for extracting valuable insights into relative stabilities and ensemble configurations. A detailed description and dataset from the \ch{NaMg(BH4)3} study are provided as an interactive tutorial included with the \texttt{crystALL} software (see \autoref{sec:softwareavailability}).

\subsection{Catalysis Materials and Molten Salts} \label{ssec:catalysisreaction}

\ch{Ru}-based catalysts are critical to enhancing manufacturing efficiency of nylon, thanks to being highly effective in synthesizing the key chemical intermediate through the benzene partial hydrogenation reaction \cite{Zhang2013}. Achieving precise control over hydrogenation selectivity involves designing diverse active site ensembles on the surfaces of intermetallic catalysts \cite{Dasgupta2022AtomicSelectivity}. The \ch{Ru-Zn} system is of particular interest to the community, but it is typically modeled strictly as FCC/HCP solid solutions \cite{Mu2025ANanoparticles} without the consideration of intermetallic compounds. To address this, \citet{Gong2025InvestigationSalts} used \texttt{crystALL} alongside \texttt{pySIPFENN} \cite{Krajewski2022ExtensibleNetworks} ML models and DFT validation to explore over 165,000 candidate configurations of \ch{Ru-Zn} intermetallic compounds, including \ch{Ru3Zn}, \ch{RuZn}, \ch{Ru4Zn9}, \ch{RuZn3}, \ch{Ru3Zn10}, \ch{RuZn6}, \ch{RuZn10}, and \ch{RuZn13}; simultaneously modeling active site ensembles of top candidates to design novel catalysts with high activity and selectivity.

Another application where accurate modeling necessitates the use of ensembles to fully capture the complexity of the underlying materials is in the study of chloride molten salts in advanced Molten Salt nuclear Reactors (MSRs) and pyroprocessing techniques to separate lanthanide fission products \cite{Gong2025ThermodynamicQuantification}, both of which are critical in the journey towards clean energy and carbon footprint reduction. In this example, \texttt{crystALL} facilitated the prediction of structures for the \ch{KLa3Cl10} and \ch{K3La5Cl18} compounds in the \ch{KCl-LaCl3} system described by \citet{Gong2025ThermodynamicQuantification}, thereby explaining their stability reported in various literature sources \cite{Seifert1985, Song1995} and allowing for further investigation using ab-initio-based methods.

\section{Summary} \label{sec:summary}

Section~\ref{sec:newstable} presents a high-level analysis of the field of computational structure prediction, where for the last several decades, two fundamental challenges have persisted: predicting (a) which compositions form compounds of interest experimentally and (b) their candidate topologies or atomic arrangements. Databases like \texttt{AFLOW} \cite{Curtarolo2012AFLOW:Discovery}, \texttt{OQMD} \cite{Saal2013MaterialsOQMD}, and Materials Project \cite{Jain2013Commentary:Innovation, Horton2025AcceleratedProject} have made invaluable contributions by systematically populating prototype libraries with diverse chemistries, creating rich repositories of both stable and metastable structures. However, some of the recent ML-driven approaches have shifted focus exclusively toward finding "new stable compounds", potentially creating a form of model collapse where iterative bias toward lowest-energy structures progressively eliminates the diversity essential for both (a) understanding real materials formed by statistical ensembles of configurations and (b) robustness on the training datasets allowing discovery of novel structures. Methods like \texttt{USPEX} \cite{Glass2006USPEXPrediction}, \texttt{AIRSS} \cite{Pickard2011iAbSearching}, and \texttt{XtalOpt} \cite{Hajinazar2026XtalOptOptimization} can in principle generate diverse polymorphs for individual compositions, but they are much too expensive to efficiently generate wide, approximately ergodic ensembles across many chemistries. This gap is critical because metastable configurations in phases often govern interesting properties (e.g., negative thermal expansion or occurrence of catalytic sites), thermodynamic models require diverse configurational data, and experimental structures frequently correspond to higher-energy configurations. \texttt{crystALL} addresses this niche by leveraging the structural diversity already present across millions of entries in \texttt{MPDD} to generate comprehensive configurational ensembles in the least biased fashion.

The beginning of Section \ref{sec:crystall} establishes the foundation of the \texttt{crystALL} approach, emphasizing how it leverages \texttt{MPDD}'s 4.9 million relaxed structures as diverse prototypes spanning configurations, including many symmetry-broken ones. This bias-minimizing approach of using all available data generates orders of magnitude more candidates than traditional prototype libraries, which can be handled thanks to the use of ML for their characterization. Section \ref{sec:crystall} also notes how users can expand this pool further through community datasets and systematic generation tools like \texttt{ATAT} \cite{vandeWalle2013EfficientStructures} and \texttt{aflow++} \cite{Oses2023Aflow++:Design}. Subsequently, \autoref{ssec:compounds} details the implementation workflow of \texttt{crystALL} where, typically, thousands of prototypes that can accommodate a given stoichiometry are used to generate candidates through all permutations of atomic substitutions, then followed by featurization in \texttt{pySIPFENN}'s \texttt{KS2022} space \cite{Krajewski2025EfficientStructures}. Such representations are then clustered through a method of choice to identify unique representative candidates picked by ML models based on the lowest energy in the cluster, i.e., the lowest sampled configuration in given basin of energy landscape, or based on match to observed properties. Finally, the top candidates can be passed to DFT calculations or matched with experiments directly. The resulting ensemble captures diverse configurations across the energy landscape, with validated structures feeding back into \texttt{MPDD} to enrich future searches.

Section \ref{ssec:compositionranges} addresses a critical gap in computational materials discovery by extending \texttt{crystALL} to handle compositional uncertainty inherent to experimental measurements. The section addresses how techniques like EDS yield uncertain, non-stoichiometric compositions, like \ch{Al_{32.1}Cu_{67.9}} $\pm 1.5~at.~\%$, that could correspond to many compounds. Traditional search techniques typically either (a) define a set of low-index anonymous formulas, risking missing prototypes with complex high-index ratios or (b) search for compositional ranges based by querying for bounds over individual chemical elements, which works well for compounds with low number of components but fails in more complex cases. \texttt{MPDD} implements sorted fractional composition queries allowing \texttt{crystALL} to query on \texttt{sortedFractionsDict} field, enabling efficient searches across anonymized and symmetry-collapsed compositional bounds; reducing query complexity $10^7$ times for quinary compounds. All prototypes within bounds undergo the standard pipeline (substitutions, featurization, clustering), with formation energies referenced to ML-predicted convex hulls for composition-corrected stability comparisons. The output provides both predicted stability and compositional distance from the nominal value, enabling informed selection for validation. 

Lastly, Section \ref{sec:casestudies} demonstrates \texttt{crystALL}'s practical use across diverse materials systems, including published validations of its effectiveness in structure discovery for experimentally observed compounds in \autoref{ssec:ndbi2}. It showcases how \texttt{crystALL} was used in \citet{Im2022ThermodynamicModeling} to find \ch{NdBi2} structure, unresolved for five decades; it identified it as a mixture of two metastable configurations through \texttt{crystALL}, while ground-state-searching methods like \texttt{USPEX} converged only to the lowest-energy structure that did not match experiments. Another validation was also achieved in \citet{Shang2021FormingJoints} for \ch{Al2Fe} intermetallics in high-pressure welding applications. Additionally, several yet-unpublished and example use cases have been included, such as identification of observed \ch{Nd-Bi-Sn} ternary compound through compositional search from \autoref{ssec:compositionranges}. \autoref{sec:novelhydrides} illustrates \texttt{crystALL}'s potential for discovering complex structures, predicting diverse \ch{NaMg(BH4)3} configurations. Lastly, \autoref{ssec:catalysisreaction} also includes examples of use over functional materials through systematic analysis of \ch{Ru-Zn} intermetallics (165,000+ candidates across several observed compounds) for benzene hydrogenation catalyst design and successful structure prediction of \ch{KLa3Cl10} and \ch{K3La5Cl18} compounds essential for molten salt reactor technologies and lanthanide separation \cite{Gong2025ThermodynamicQuantification}.

Overall, the \texttt{crystALL} approach represents a shift in computational materials discovery to fill an interesting niche - from focusing on singular ground states to embracing the full configurational complexity, metastability, and uncertainty that defines real materials at finite temperatures. It conceptually bridges the gap between idealized computational predictions and  experimental observations. By leveraging millions of existing configurations relaxed for hundreds of thousands of different chemical compositions, it forms a minimally-biased and ensemble-focused framework. The \texttt{crystALL} approach generates intermediate results, which do not find "the answer" but generate comprehensive landscapes of possibilities to explain experimental results and empower novel materials discovery.

\section{Software and Data Availability}
\label{sec:softwareavailability}

The \texttt{crystALL}  software described in this work has been published as free open-source software (FOSS) under the MIT license. It is available through the GitHub repository at \href{https://github.com/PhasesResearchLab/crystALL}{github.com/PhasesResearchLab/crystALL} and selected versions are archived in Zenodo under all-versions DOI \href{https://doi.org/10.5281/zenodo.23872006}{10.5281/zenodo.23872006}, including \texttt{v0.4.0} used in \citet{Im2022ThermodynamicModeling} and \citet{Shang2021FormingJoints}, and the latest stable release \texttt{v0.5.4} which is also attached to this publication.

End-users can also obtain it from the \texttt{PyPI} package repository under the name \href{https://pypi.org/project/crystall-search}{\texttt{crystall-search}}.

Additionally, the upcoming \texttt{v0.6.0} version featuring the \texttt{MLIPAdapter} relaxation architecture covering 24 MLIPs, described in \autoref{app:mlips}, is currently available for cloning and testing via the \texttt{v0.6.0} development branch on the GitHub repository.

Data used within this study and the past studies employing \texttt{crystALL} has been obtained from \texttt{MPDD}, which can be accessed via an open-access \texttt{OPTIMADE} endpoint at \href{https://optimade.mpdd.org}{optimade.mpdd.org} indexed by the consortium, or via a closed-access API authors can share upon reasonable request.

\section*{Contributions CRediT}
\textbf{Adam M. Krajewski:} Conceptualization, Methodology, Software, Writing - Original Draft, Data curation, Validation, Visualization; \textbf{Ricardo Amaral:} Methodology, Software, Investigation, Writing - Original Draft, Visualization;
\textbf{Jonathan W. Siegel:} Software, Validation, Writing - Review \& Editing;
\textbf{Luke A. Myers:} Methodology, Software, Writing - Original Draft;
\textbf{Rushi Gong}: Investigation, Writing - Original Draft;
\textbf{Shun-Li Shang:} Investigation, Writing - Review \& Editing;
\textbf{Zi-Kui Liu:} Funding acquisition, Supervision, Writing - Review \& Editing, Resources.

\section*{Acknowledgments}

This work has been funded through grants: U.S. National Science Foundation - Pathways to Enable Open-Source Ecosystems (POSE) with Award No. \textbf{FAIN-2229690}, NSF Grant \textbf{DMS-2424305}, DOE Advanced Research Projects Agency-Energy (ARPA-E) with Award No. \textbf{DE-AR0001435}, and DOE Nuclear Energy University Programs (NEUP) with Award No. \textbf{DE-NE0009288}.

We would like to thank \textbf{Dr. Hongyeun Kim} at PSU for valuable feedback on the development of the compositional search method, and \textbf{Dr. Hojong Kim}’s research group at PSU for conducting experiments in the Bi-Nd and Bi-Nd-Sn systems, which inspired the initial method development in 2020.

Additionally, Ricardo Amaral gratefully acknowledges financial support from the \textbf{Cuomo Foundation} and the \textbf{Intergovernmental Panel on Climate Change (IPCC)}. The contents of this document are solely the responsibility of the authors and under no circumstances may be considered as reflecting the position of the Cuomo Foundation and/or the IPCC.

\printbibliography 

\newpage

\appendix
\onecolumn

\addtocontents{toc}{\protect\setcounter{tocdepth}{0}}

\titleformat{\section}
  {\normalfont\Large\bfseries}{Appendix \thesection.}{1em}{}

\counterwithin{figure}{section}
\counterwithin{table}{section}
\renewcommand{\thefigure}{\thesection\arabic{figure}}
\renewcommand{\thetable}{\thesection\arabic{table}}

\section{\texttt{crystALL} clustering of structures} \label{app:crystallclustering}

\begin{figure}[!hb]
\centering
\includegraphics[width=0.94\textwidth]{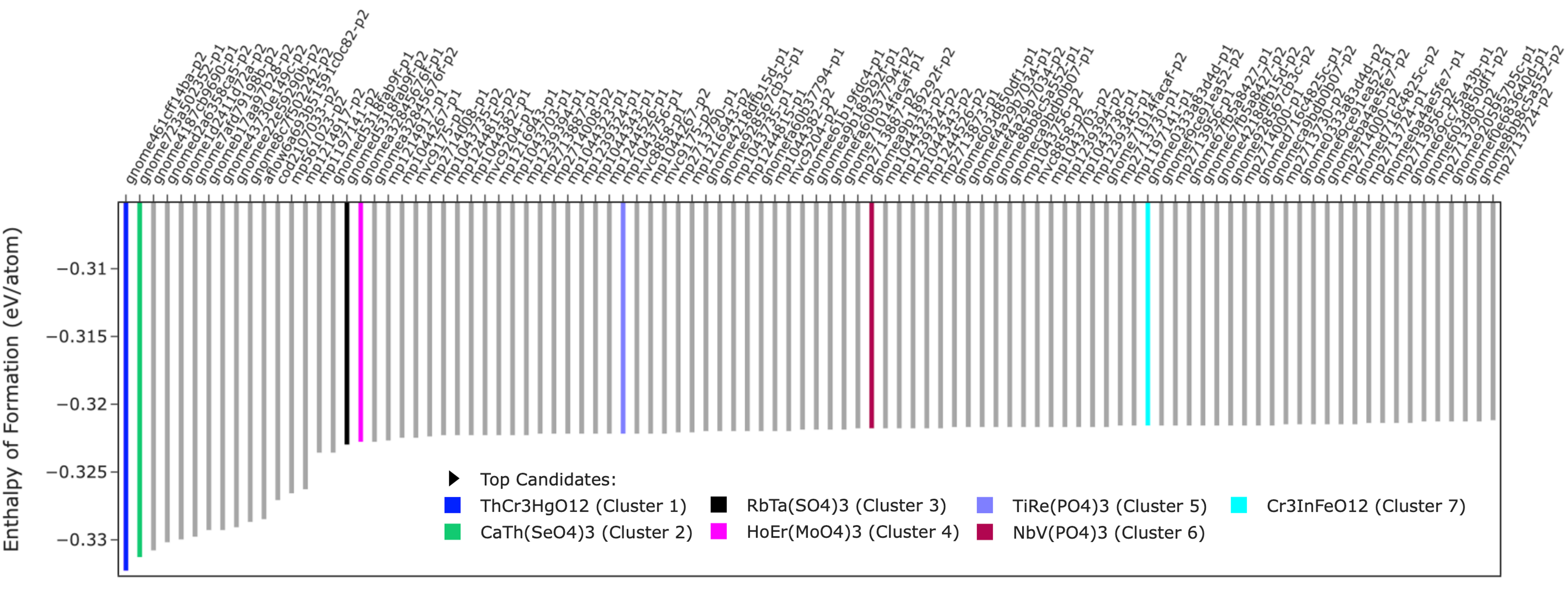}
\caption{Bar chart with the 100 most stable prototypes and 7 top candidates per cluster identified by \texttt{crystALL} for the \ch{NaMg(BH4)3} composition. Top candidate structures identified on the graph are based on the prototypes: \ch{ThCr3HgO12} (gnome461cff14ba), \ch{CaTh(SeO4)3} (gnome723a502352), \ch{RbTa(SO4)3} (gnomed5318fab9f), \ch{HoEr(MoO4)3} (gnomea32845676f), \ch{TiRe(PO4)3} (mp1043756), \ch{NbV(PO4)3} (gnomea9b189292f), and \ch{Cr3InFeO12} (gnome0333a83d4d).}
\label{fig:hydrides_chart}
\end{figure}

\begin{figure}[H]
    \centering
    \includegraphics[width=0.94\textwidth]{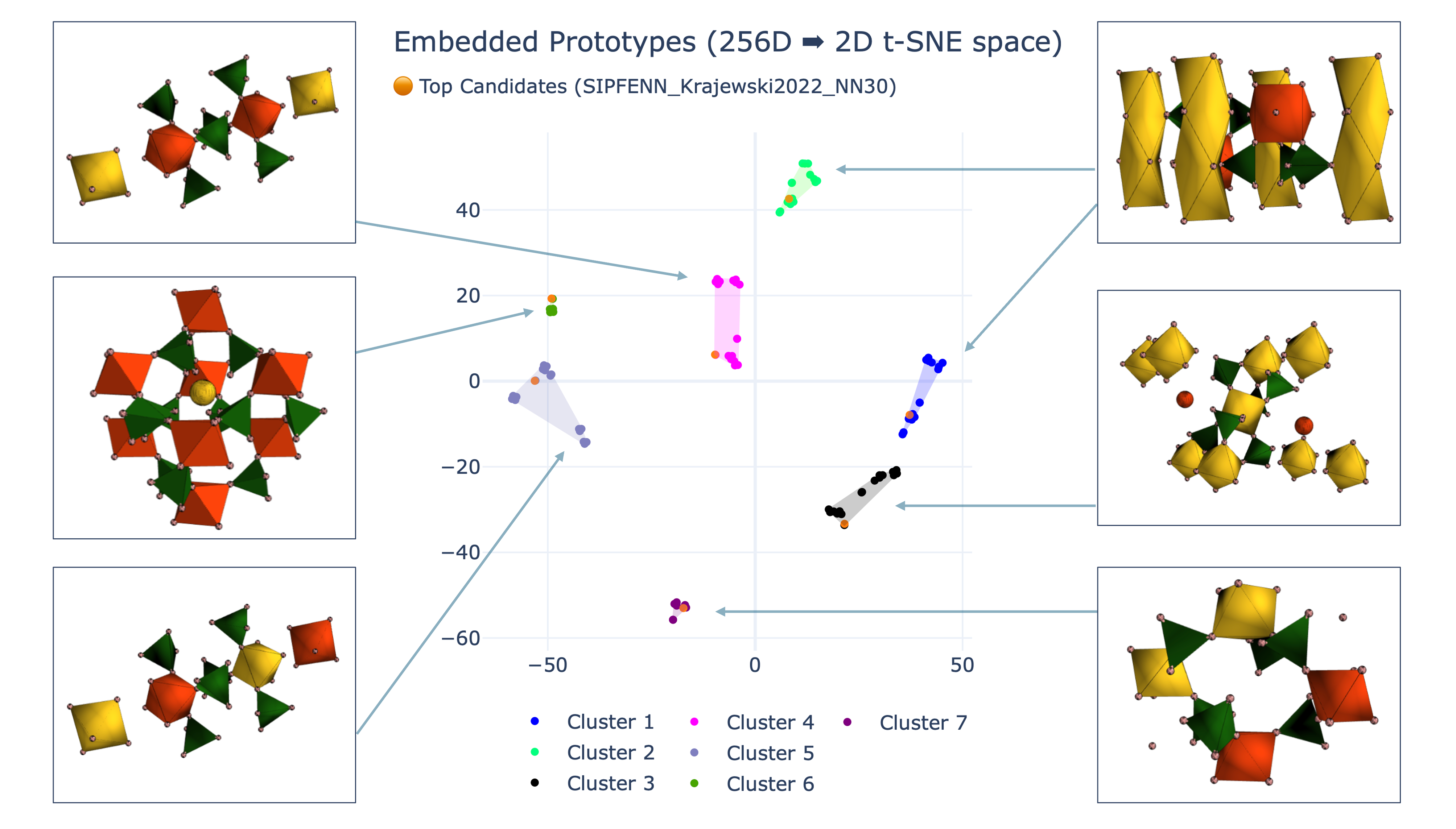}
    \caption{2D tSNE and k-means clustering of the 100 most stable \ch{NaMg(BH4)3} structures predicted by \texttt{crystALL}, revealing seven distinct clusters: hexagonal (clusters 1 and 2), triclinic (cluster 3), trigonal (clusters 4, 5, and 7), and perovskite-type (cluster 6) symmetries.}
    \label{fig:hydrides_cluster}
\end{figure}

To illustrate the clustering workflow implemented in \texttt{crystALL}, we present results for the 100 most stable \ch{NaMg(BH4)3} structures, as shown in Figure \ref{fig:hydrides_chart}. The clustering through tSNE and k-means identified seven distinct clusters (Fig. \ref{fig:hydrides_cluster}), effectively separating structures with fundamentally different atomic arrangements while grouping those with similar local environments. Specifically, clusters 1 and 2 primarily contained hexagonal structures, cluster 3 grouped triclinic structures, clusters 4, 5 and 7, trigonal structures, and cluster 6 contained perovskite-type structures. 

\begin{figure}[H]
    \centering
    \includegraphics[width=0.94\textwidth]{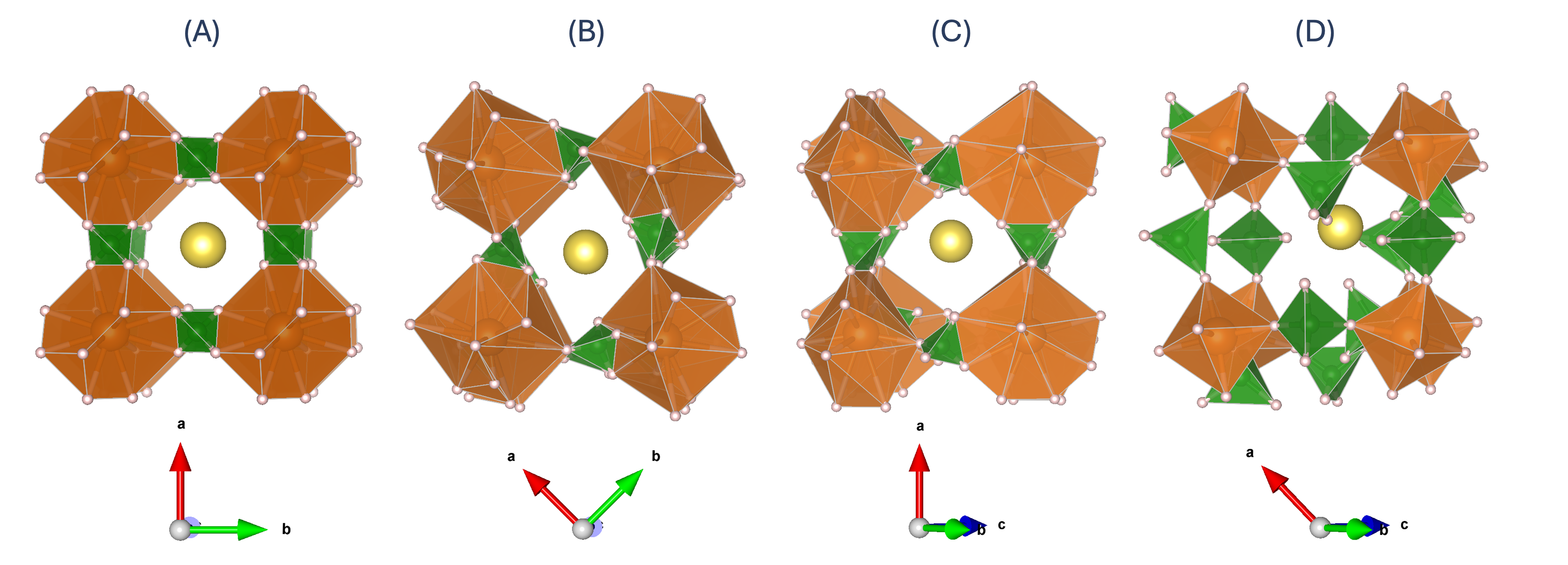}
    \caption{Side-by-side comparison of perovskite-type \ch{NaMg(BH4)3} structures based on: (A) \ch{NH4Ca(BH4)3} cubic prototype with edge-sharing octahedra (\citet{Schouwink2015}); (B) \ch{CsSr(BH4)3} prototype with edge- and face-sharing octahedra (\citet{Moller2016}); (C) \ch{RbSr(BH4)3} prototype with edge- and corner-sharing octahedra (\citet{Moller2016}); and (D) the most stable candidate from cluster 6 predicted by \texttt{crystALL}, based on the \ch{NbV(PO4)3} prototype (GNoME dataset entry a9b189292f), featuring unique corner-sharing coordination not observed in previous reports.}
    \label{fig:hydrides_perovskites}
\end{figure}

We compared the most stable \ch{NaMg(BH4)3} prototypes predicted by \texttt{crystALL} with previously reported borohydride perovskite structures from the literature, as shown side-by-side in Figure \ref{fig:hydrides_perovskites}. The literature examples include both edge-sharing and face-sharing configurations described by \citet{Schouwink2015} and \citet{Moller2016}. In contrast, the most stable candidate from cluster 6, derived from the \ch{NbV(PO4)3} prototype (GNoME dataset entry a9b189292f), adopts a highly distorted perovskite structure with unique corner-sharing coordination of the \ch{BH4-} anions, distinguishing it from previously reported borohydride frameworks.

\section{Machine Learning Interatomic Potentials (MLIPs) Supported in v0.6.0} \label{app:mlips}

The upcoming \texttt{crystALL} \texttt{v0.6.0} release (currently accessible via the \texttt{v0.6.0} branch on GitHub) introduces a fully modular architecture driven by the \texttt{MLIPAdapter} registry, enabling automated structure relaxations and multi-property predictions across diverse machine-learned interatomic potentials. Each potential manages its own execution micro-environment with support for CPU and CUDA acceleration, full cell/volume optimization, and optional piping into downstream static property models. 

A summary of the machine-learned interatomic potentials integrated into \texttt{crystALL} \texttt{v0.6.0} is given in Table~\ref{tab:mlip_models}. In the table, the ``Checkpoint Pinned'' column refers to the default model checkpoint variant adopted. However, users can easily access other checkpoint variants within each model family (such as models with different parameter sizes, training datasets, or dispersion corrections) by updating the \texttt{MLIPAdapter} configuration file or supplying custom parameters at initialization. Newly released potentials can be added in a similar fashion by updating the same configuration file, and newly vetted models will continuously be featured in upcoming minor releases. The release column distinguishes between models distributed via official PyPI packages and those installed from source repositories. Official PyPI releases are preferred for their stable dependencies and guaranteed forward compatibility, whereas ``From Source'' configurations are provisional and should be replaced as official packages become available.

Unlike static predictors such as \texttt{pySIPFENN} and \texttt{ALIGNN}, which evaluate fixed geometries, MLIPs iteratively optimize atomic positions and cell parameters. In \texttt{crystALL} \texttt{v0.6.0}, multiple MLIPs can be evaluated as separate objectives for the same candidate structures, with each potential relaxing independently to prevent coordinate conflicts. During relaxation, adapters extract converged observables such as relaxed total and formation energies, residual forces, stresses, cell volume, and the optimized geometry. Furthermore, static adapters can be piped into any MLIP to evaluate the relaxed structures directly, pairing the chosen potential with multiple downstream property models.


\begin{table*}[t]
  \centering
  \footnotesize
   \caption{Summary of machine-learned interatomic potentials (MLIPs) supported in \texttt{crystALL} \texttt{v0.6.0}. For each model family, ``Checkpoint Pinned'' denotes the default pre-trained variant loaded by the adapter (user-configurable). Release indicates distribution through official PyPI packages or installation from source repositories. Hardware lists supported execution devices (CPU and CUDA).}
  \label{tab:mlip_models}
  \begin{tabular*}{\textwidth}{@{\extracolsep{\fill}} l l l c l @{}}
    \hline
    Model Family & Checkpoint Pinned & Hardware & Distribution & Citation \\
    \hline
    \textbf{DeePMD}       & DPA-4.0.1-Pro-MPtrj           & CPU, CUDA & From source & \cite{wang2018deepmd} \\
    \textbf{CHGNet}       & 0.3.0                         & CPU, CUDA & PyPI        & \cite{Deng2023CHGNetModelling} \\
    \textbf{TensorNet}    & PES-MatPES-PBE-2025.2         & CPU, CUDA & PyPI        & \cite{simeon2023tensornet} \\
    \textbf{ORB}          & v3-omat:float32-high          & CPU, CUDA & PyPI        & \cite{Rhodes2025Orb-v3:Scale} \\
    \textbf{ORB-D3}       & v3-omat:float32-high:PBE:BJ   & CPU, CUDA & PyPI        & \cite{Rhodes2025Orb-v3:Scale} \\
    \textbf{Allegro}      & OAM-L:0.1                     & CUDA      & From source & \cite{musaelian2023allegro} \\
    \textbf{AlphaNet}     & alex-0410:oma                 & CPU, CUDA & From source & \cite{yin2025alphanet} \\
    \textbf{Eqnorm}       & eqnorm-mptrj                  & CPU, CUDA & From source & \cite{chen2025eqnorm} \\
    \textbf{eSEN}         & eSEN-30M-OAM                  & CPU, CUDA & From source & \cite{fu2025esen} \\
    \textbf{MACE}         & mh-1:omat-pbe                 & CPU, CUDA & PyPI        & \cite{batatia2022mace} \\
    \textbf{MatterSim}    & v1.0.0-5M                     & CPU, CUDA & PyPI        & \cite{Yang2024MatterSim:Pressures} \\
    \textbf{NequIP}       & OAM-XL:0.1                    & CUDA      & PyPI        & \cite{batzner2022nequip} \\
    \textbf{Nequix}       & oam-1                         & CPU, CUDA & From source & \cite{koker2025nequix} \\
    \textbf{PolarMACE}    & polar-1-l                     & CPU, CUDA & From source & \cite{batatia2026polarmace} \\
    \textbf{QET}          & PES-MatPES-PBE-2025.2         & CPU, CUDA & From source & \cite{ko2025qet} \\
    \textbf{SevenNet}     & omni-i12:omat24               & CPU, CUDA & PyPI        & \cite{park2024sevennet} \\
    \textbf{UPET}         & oam-xl                        & CPU, CUDA & PyPI        & \cite{bigi2026upet, Malosso2026High-qualityLearning} \\
    \textbf{EquiformerV3} & omat24-mptrj-salex-gradient   & CPU, CUDA & From source & \cite{liao2026equiformerv3} \\
    \textbf{MatRIS}       & 10m-oam:efsm                  & CPU, CUDA & From source & \cite{zhou2026matris} \\
    \textbf{UMA}          & s-1p2:omat                    & CPU, CUDA & PyPI        & \cite{wood2025uma} \\
    \textbf{GRACE}        & 2L-SMAX-OMAT-large            & CPU, CUDA & PyPI        & \cite{bochkarev2024grace} \\
    \textbf{TACE}         & OAM-L                         & CPU, CUDA & From source & \cite{xu2025tace} \\
    \textbf{HIENet}       & 0                             & CPU, CUDA & From source & \cite{yan2025hienet} \\
    \textbf{OSAKA}        & osaka26:small                 & CPU, CUDA & PyPI        & \cite{kuroda2026osaka} \\
    \hline
    \textbf{Total}        & \textbf{24 Models}            &           &             & \\
    \hline
  \end{tabular*}
\end{table*}

\end{document}